# Simulating Capacitances to Silicon Quantum Dots: Breakdown of the Parallel Plate Capacitor Model

Ted Thorbeck, Akira Fujiwara, and Neil M. Zimmerman

*Abstract*— Many electrical applications of quantum dots rely on capacitively coupled gates; therefore, to make reliable devices we need those gate capacitances to be predictable and reproducible. We demonstrate in silicon nanowire quantum dots that gate capacitances are reproducible to within 10% for nominally identical devices. We demonstrate the experimentally that gate capacitances scale with device dimensions. We also demonstrate that a capacitance simulator can be used to predict measured gate capacitances to within 20%. A simple parallel plate capacitor model can be used to predict how the capacitances change with device dimensions; however, the parallel plate capacitor model fails for the smallest devices because the capacitances are dominated by fringing fields. We show how the capacitances due to fringing fields can be quickly estimated.

Q UANTUM dots (QDs) with capacitively coupled gates have enabled the manipulation of individual electrons, which leads to many attractive applications. For example, the single electron transistor (SET)[1] consists of a QD tunnel coupled to a source and drain, with a capacitively coupled gate that controls the average charge on the QD. The SET, in which the current can change by orders of magnitude when the charge on a nearby gate changes by less than the charge of an electron, is the world's most sensitive charge electrometer[2]. By pulsing the gates on a multiple QD device, electrons can be pumped through the QDs to define either the ampere or the farad[3,4]. SET based single electron logic and memory have been proposed as the ultimate in classical electronic scaling[5,6]. Going beyond classical electronics, both the charge and the spin of a single electron in a QD have been proposed as the basis for a quantum computer[7,8] The potential for single electron QDs is enormous, but limitations in the reproducibility and predictability of QDs have limited these devices[6]. In particular, some of these applications are impossible without reproducible gate capacitances. For example, a useful QD quantum computer or multiple-QD pump would need many devices operating in parallel with the same gate voltages pulses, which would require reproducible and predictable capacitances.

In this letter, we show that the gate capacitances in our devices, QDs in a silicon nanowire, are reproducible for devices with the same lithographic parameters, and are predictable over a wide range of capacitances for devices with different dimensions. For nominally identical devices, the measured gate capacitances are reproducible to about within 10%. In addition, using a capacitance simulator we can predict the gate capacitances to within 20%. In the capacitance simulation, we used only the parameters used during device fabrication, and no adjustable parameters, to determine the device dimensions.

For a single set of lithographically identical devices, we have previously shown that gate capacitances are reproducible.[9] Additionally it has been shown that gate capacitances can scale with lithographic parameters[10–12]. In this work we will numerically describe the reproducibility of gate capacitances. Other groups have demonstrated the ability to simulate gate capacitance;[13,14] however, most have done so for only one device, and some rely on several fitting parameters to match the experimental capacitances. We demonstrate capacitance simulations for a wide variety of device dimensions with no fitting parameters. Some previous work[11,12] has relied upon a parallel plate capacitance model to predict capacitances. In this approximation only field lines from the gate straight to the QD are included. In contrast, for our smallest devices the capacitances are instead dominated by fringing field lines from the gate to the QD.

Our devices[9] (Fig. 1(a)) consist of a lightly p-doped or undoped nanowire surrounded by two gate layers. The nanowire is mesa-etched into an SOI (silicon on insulator) wafer. Devices with a wide range of dimensions are presented in this paper. The minimum (maximum) dimensions of the nanowire: 17 (21) nm thick ($t_{si}$), 10 (30) nm wide ($w_{si}$), 300 (510) nm long. Thermally grown $SiO_2$ isolates the nanowire from the gates ($t_{ox,1}$ = 20 nm). The lower gate layer consists of This work was supported in part by the Laboratory for Physical Sciences (EAO93195).

Ted Thorbeck is with the National Institute of Standards and Technology, Gaithersburg, MD 20899 USA and the Joint Quantum Institute at the University of Maryland, College Park, MD 20740 USA. (phone; 301-957-4133; e-mail: tedt@nist.gov).

Akira Fujiwara is with NTT Basic Research Laboratories, NTT Corporation, 3-1 Morinosato Wakamiya, Atsugi, Kanagawa 243-0198, Japan (e-mail: fujiwara,akira@lab.ntt.co.jp).

Neil M. Zimmerman is with the National Institute for Standards and Technology, Gaithersburg, MD 20899 USA (phone: 301-975-4270; URL: ftp://ftp.nist.gov/pub/physics/neilz/papers.html).

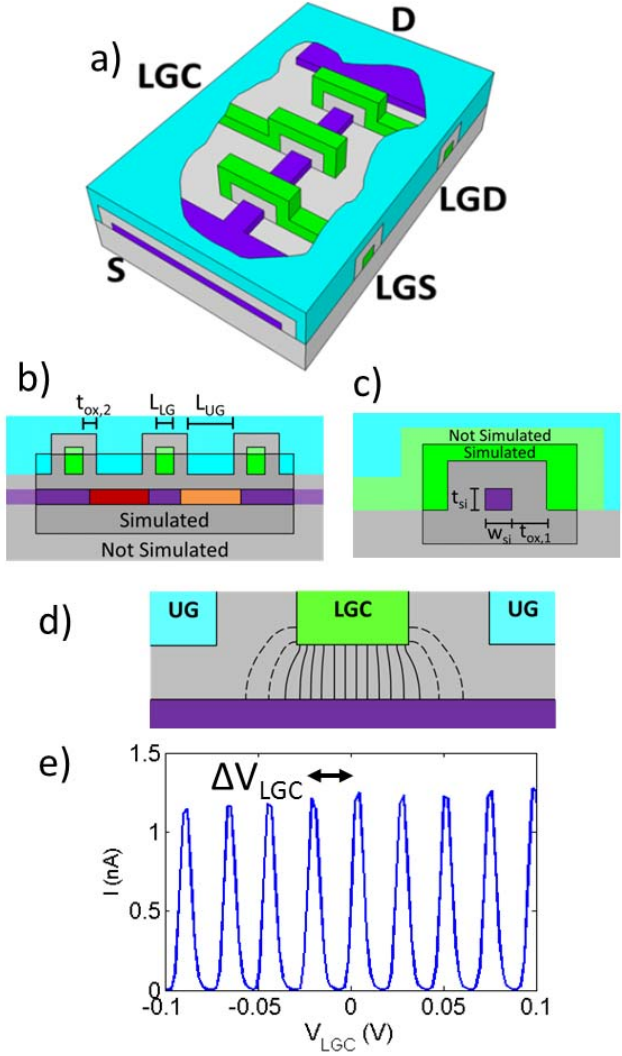

Fig. 1. (Color Online) (a) A cutaway schematic of the device, with the nanowire in purple, the lower gates in green and the upper gate in blue. (b) & (c) Cross sections of the device showing dimensions and identifying what was simulated in FASTCap. The red and orange boxes correspond to the source and drain half QDs. (d) Schematic of electric field lines from LGC to full island, the direct field lines (corresponding the parallel plate capacitance model) are solid, and the fringing electric field lines are dashed. Representation of the electric field lines was cutoff when the strength reached half of the peak strength. (e) Coulomb blockade oscillations with $C_{LGC}$ = 6.7 aF for a device with $w_{si}$ = 30 nm, $t_{si}$ = 21 nm, $L_{LG}$ = 10 nm and $L_{UG}$ = 40nm.

three heavily doped poly-silicon gates that wrap around three sides of the nanowire like a tri-gate-FET (field effect transistor). The lower gates are 10 (40) nm long ($L_{LG}$) and spaced by 70 (130) nm (S). More $SiO_2$ separates the lower gate layer from the global upper gate ($t_{ox,2}$ = 30 nm). The length of the upper gate ($L_{UG} = S - 2t_{ox,2}$) ranges between 40 nm and 100 nm, and corresponds to the length the lower gates fills in between the two lower gates and is closest to the nanowire.

A positive voltage on the upper gate inverts the nanowire. Negative voltages on the lower gates deplete the nanowire, eventually forming tunnel barriers. A QD forms when any two lower gates create tunnel barriers. The outer two lower gates (LGS and LGD) define the full QD. The source (drain) half QD is defined by LGC and LGS (LGD) (Fig. 1(b)). The QD is used as the island of an SET, so that when the voltage on any capacitively coupled gate is ramped, the current through the SET oscillates periodically (Fig. 1(e)). The period of the peaks in gate voltage is inversely proportional to the gate capacitance, $(C_G = |e|/\Delta V_G)^1$. We used this formula to measure the capacitance from each of the four gates to each of the three QDs (one full and two half QDs). We have previously shown that in this geometry the gate capacitances are not sensitive to gate voltages[9].

We used FASTCap[15,16], an electromagnetic field solver, to simulate the gate capacitances. The simulation was based on our calculation of the device dimensions as fabricated. This was determined using only the following parameters: $t_{si}$, $w_{si}$, S, $L_{LG}$, $t_{ox,1}$, and $t_{ox,2}$. We do not use any fitting parameters in the simulation. In the simulator, we assumed that the QD (approximated as a perfect metal) occupies all four sides of the nanowire and that the QD terminates 15 nm ($T_{ox,2}/2$) before the near end of the lower gate causing the tunnel barrier. The geometry that we simulated is shown in Fig. 1(b) and (c). All structures in the simulator were terminated 50 nm away from the QDs because conductors and dielectrics further away had little impact on the simulated capacitances but increased the runtime of the simulation.

Table I contains the full set of measured gate capacitances (UG, LGS, LGC, LGD) to each of the three QDs (full, source half, and drain half) for two lithographically identical devices as well as the simulated capacitances. There are several qualitative dependences which make intuitive sense. 1) For the full QD, the barrier gates (LGS and LGD) have similar capacitances to the QD and LGC has a larger capacitance to the QD, as expected. 2) For the source half QD, LGS and LGC have similar capacitances to the QD whereas LGD has a much smaller capacitance because it is far away and screened by the other gates. 3) The capacitance from UG to the full QD is roughly the sum of the capacitances from UG to the source

TABLE I
MEASURED AND SIMULATED CAPACITANCES

| Cap (aF) | UG | | | LGS | | | LGC | | | LGD | | |
|---|---|---|---|---|---|---|---|---|---|---|---|---|
| | Dev. 1 | Dev. 2 | Sim. | Dev. 1 | Dev. 2 | Sim. | Dev. 1 | Dev. 2 | Sim. | Dev. 1 | Dev. 2 | Sim. |
| Full Island | 22 | 22 | 25 | 2.7 | 3.0 | 2.6 | 6.2 | 6.0 | 6.4 | 2.5 | 2.8 | 2.6 |
| Source Half | 11 | 10 | 12 | 2.3 | 2.8 | 2.6 | 2.8 | 2.6 | 2.6 | 0.1 | 0.1 | 0.1 |
| Drain Half | 11 | 11 | 12 | 0.1 | 0.1 | 0.1 | 3.1 | 2.8 | 2.6 | 2.4 | 2.5 | 2.6 |

Measured gate capacitances for device 1 and device 2 (nominally identical) and simulated capacitances for each of the four gates to each of the three QDs. For comparison to Figs. 2 & 3 the area of the full QD under the lower gate is 940 $nm^2$, and the area of the half QD under the upper gate is 3760 $nm^2$.

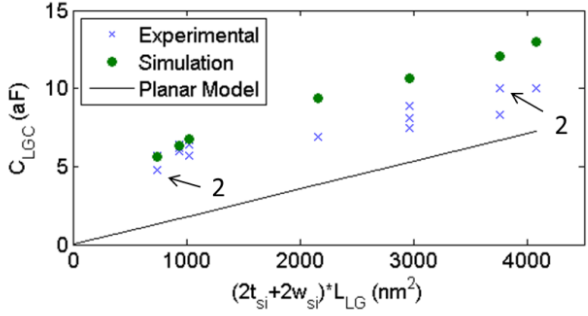# 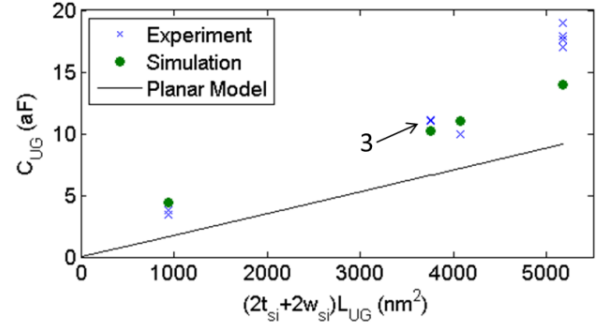

Fig. 2. Measured capacitances from LGC to the full island (blue x's), simulated capacitances (green circles) and capacitances calculated from the parallel capacitance model (black line), which show reproducibility of measured capacitances and that the simulation predicts measured capacitances much better than the parallel plate model. The horizontal axis represents the area of the nanowire directly below the lower gate, $(2t_{si}+2w_{si})L_{LG}$, based on the parameters used in lithography. Arrows with a number correspond to multiple data points on top of each other. The uncertainties are not shown, but are typically smaller than the plotting symbols.

Fig. 3. Experimental and simulated capacitances from UG to either source or drain half QD parameterized for the area of the nanowire closest to the upper gate, $(2t_{si}+2w_{si})L_{UG}$. Symbols are the same as in Fig 2.

and drain half QDs. To quantitatively describe the reproducibility in Table 1, we define deviation as the difference between an individual data point and the average divided by the average. All the deviations in Table I are less than 15% and the average deviation is 6%. This demonstrates the ability to reproduce capacitances to lithographically identical devices.

Figure 2 shows the measured and simulated values for the capacitance from LGC to the full QD. To compare devices of different lithographic parameters, we combine the thickness of the nanowire ($t_{si}$), width of the nanowire ($w_{si}$) and length of the lower gate ($L_{LG}$) into a single parameter that describes the surface area of the nanowire directly below the gate, $(2t_{si}+2w_{si})L_{LG}$. Three points stand out when comparing the measured and simulated capacitances in Fig. 2. 1) For devices with the same surface area parameter, the maximum deviation is 9% and the average deviation is 7%. This demonstrates the reproducibility of the gate capacitance for nominally identical devices over a wide range of device dimensions. 2) The spread of capacitances for a single surface area is smaller than the range of capacitances for different surface areas; this means that the lithographic parameters are controlling the gate capacitance. 3) The simulated capacitances accurately predict the measured capacitances. The average deviation of the simulation (absolute value of the difference between the average measured and simulated capacitance divided by the simulated capacitance) is 17%. We note that, for larger devices, the capacitance simulator does tend to predict a larger $C_{LGC}$ than measured. One possible explanation is that the thermal oxidation rate depends on the stress in the silicon; in a larger device less stress might build up than expected, allowing for more silicon to be consumed by the oxide therefore making the capacitance smaller than we calculated (resulting in a larger $t_{ox,2}$, and smaller $L_{LG}$)[10].

Figure 2 also shows the breakdown of the parallel plate capacitance model. Although the parallel plate capacitance model captures the slope of the experimentally measured $C_{LGC}$, it predicts an intercept of zero whereas the measured $C_{LGC}$ has an intercept of 4.7 aF. For the smallest devices studied, the average measured $C_{LGC}$ is 5.1 aF; the capacitance simulator predicts 5.6 aF, a 10% error, whereas the parallel plate model predicts a capacitance of 1.0 aF. For smaller devices this discrepancy will increase. This breakdown is not surprising; we only expect the parallel plate capacitance model to work if the plates of the capacitor are much larger than their separation. In contrast, for our smallest devices fringing electric field lines dominate the capacitance between gate and dot (Fig. 1(d)). A simple way to estimate the fringing fields is to note that the fringing fields should fall off on a length scale about equal to the separation of the gate and dot, in our case $t_{1,ox}/2$. Therefore a corrected parallel plate capacitance estimate would look like: $C_{LGC} = \varepsilon(2t_{si}+w_{si})(L_{LG}+t_{1,ox})/t_{1,ox}$. For our smallest devices this corrected parallel plate capacitor predicts $C_{LGC} = 3.9$ aF, which favorable compares to our average measured $C_{LGC} = 5.1$ aF.

Figure 3 shows both the simulated and measured capacitance from UG to either the source or drain half QD (which by symmetry should have identical capacitances to UG). Here we parameterize the area as $(2t_{si}+2w_{si})L_{UG}$, where $L_{UG}$ is the length over which the upper gate is closest to the QDs (Fig. 1(b)). This plot makes the same three points the previous plot did. 1) The spread of capacitances for a single surface area parameter is small. The average deviation is 5% and the maximum deviation is 6%. 2) The lithographic parameters control the gate capacitances. 3) The simulation predicts the measured gate capacitances; the average simulated deviation is 14%. Figure 3 also show how for the smallest devices the capacitance simulator does a much better job of predicting gate capacitances than the uncorrected parallel plate capacitance model. The parallel plate capacitance model again underestimates the capacitance. For the smallest measured devices shown in Fig. 3, the average measured capacitance is $C_{UG} = 3.6$ aF. The uncorrected parallel plate capacitor estimates $C_{UG} = 1.6$ aF. Once we add the fringing fields to the corrected parallel plate capacitor we estimate a capacitance of 4.0 aF. This again demonstrates that for our smallest devices the capacitance is dominated by fringing electric fields. To explain the difference between the simulated and measured $C_{UG}$ for the largest devices in Fig. 3 we suspect that differences in wire length affects thickness of the oxide, $t_{ox,1}$.

For both Figs. 2 & 3 the parallel plate capacitance model did a better job of predicting measured capacitances when the full four-sided perimeter of the wire ($2t_{si}+2w_{si}$) was used in calculating the surface area than when only the three-sides adjacent to the gates ($2t_{si}+w_{si}$) was used. This suggests that although we have a tri-gate-FET, conduction is occurring along all four surfaces of the wire.

In contrast to our success in simulating gate capacitances, the simulation was unable to match the measured barrier capacitances (the capacitance from the source or drain to the QD). We have previously shown that the barrier capacitance is a function of gate voltage[17], with a minimum of 15 aF and a maximum of 50 aF (for a device $w_{si}$ = 30 nm, $t_{si}$ = 10), whereas our simulation predicted 4 aF. To explain the gate voltage dependence of the barrier capacitances, either the length of the barrier or the dielectric constant ($\varepsilon$) must be a function of gate voltage. If the length of the barrier were a function of the gate voltage, then the length of the QD and hence the gate capacitances to the QD would be a function of the gate voltage, which we do not observe. A more likely explanation is related to $\varepsilon$; as $V_{LG}$ becomes less negative the nanowire carrier density ($\rho$) increases, and the silicon changes from insulating to semiconducting. Also, as $\rho$ increases, we expect that $\varepsilon$ increases from its textbook value, which corresponds to low $\rho$. Such a capacitance enhancement has been observed in doped[18] and undoped[17] silicon tunnel barriers, but no simulation has succeeded in predicting measured capacitances. Significant additional work is needed to include the gate voltage dependence in the simulation. To generalize from our experience, capacitors with a chemically defined insulator (like our gate capacitances) can be successfully simulated, whereas capacitors with an electrically defined insulator (like our barrier capacitances) cannot be successfully simulated.

In summary, the combination of Table I and Figs. 2 and 3 demonstrate experimental control over gate capacitances using the fabrication parameters. We showed that gate capacitances are typically reproducible to within 10% for lithographically identical devices. We showed how the often used parallel plate capacitance model breaks down for small devices because the capacitance is dominated by fringing electric fields. We showed how the parallel plate model can be improved by estimating the fringing capacitance. We predicted the gate capacitances in a capacitance simulator to within 20% using only lithographic parameters and no fitting parameters. Having a well calibrated capacitance simulator for these devices allows us to apply it to other ends, such as determining the location of QDs not defined by the gates which frequently occur in these and other silicon devices[19].

We would like to acknowledge helpful conversations with Michael Stewart, Panu Koppinen, Russell Lake, Josh Pomeroy and Harold Stalford.